\documentclass[acmsmall]{acmart}
\usepackage{multirow}
\usepackage{fancyhdr}

\AtBeginDocument{%
  \providecommand\BibTeX{{%
    \normalfont B\kern-0.5em{\scshape i\kern-0.25em b}\kern-0.8em\TeX}}}

\renewcommand\footnotetextcopyrightpermission[1]{} % removes footnote with conference information in first column
\setcopyright{acmcopyright}
\copyrightyear{2021}
\acmYear{2021}
\acmDOI{01.2345/6789012.3456789}

\acmJournal{JACM}
\acmVolume{1}
\acmNumber{1}
\acmArticle{1}
\acmMonth{2}

\begin{document}

\title[]{A Survey of Blockchain Data Management Systems}

\author{Qian Wei}
%\email{webmaster@marysville-ohio.com}
\affiliation{%
  \institution{Shandong University}
  %\streetaddress{72 Binhai Rd}
  %\city{Qingdao}
  %\state{Qingdao shi}
  \country{China}}
  %\postcode{266237}}
  \email{weiqian1212@mail.sdu.edu.cn}

\author{Bingzhe Li}
\affiliation{%
  \institution{Oklahoma State University}
  %\streetaddress{}
  %\city{Oklahoma}
  %\state{Beijing Shi}
  \country{USA}}
  \email{bingzhe.li@okstate.edu}

\author{Wanli Chang}
\affiliation{%
  \institution{University of York}
  %\streetaddress{}
  %\city{York}
  %\state{Beijing Shi}
  \country{UK}
  }
  \email{wanli.chang@york.ac.uk}

\author{Zhiping Jia}
\affiliation{%
  \institution{Shandong University}
  %\streetaddress{72 Binhai Rd}
  %\city{Jimo Qu}
  %\state{Qingdao shi}
  \country{China}}
  %\postcode{266237}}
  \email{jzp@sdu.edu.cn}

\author{Zhaoyan Shen}
\affiliation{%
  \institution{Shandong University}
% \streetaddress{72 Binhai Rd}
 % \city{Jimo Qu}
  %\state{Qingdao shi}
  \country{China}}
  %\postcode{266237}}
  \email{shenzhaoyan@sdu.edu.cn}

\author{Zili Shao}
\affiliation{%
  \institution{The Chinese University of Hong Kong}
  %\streetaddress{}
  %\city{Hong Kong}
  %\state{Beijing Shi}
  \country{China}}
  \email{zilishao@cuhk.edu.hk}

\renewcommand{\shortauthors}{}

\begin{abstract}

Blockchain has been widely deployed in various sectors, such as finance, education, and public services. Since blockchain runs as an immutable distributed ledger, it has decentralized mechanisms with persistency, anonymity, and auditability, where transactions are jointly performed through cryptocurrency-based consensus algorithms by worldwide distributed nodes. There have been many survey papers reviewing the blockchain technologies from different perspectives, e.g., digital currencies, consensus algorithms, and smart contracts. However, none of them have focused on the blockchain data management systems. To fill in this gap, we have conducted a comprehensive survey on the data management systems, based on three typical types of blockchain, i.e., standard blockchain, hybrid blockchain, and DAG (Directed Acyclic Graph)-based blockchain. We categorize their data management mechanisms into three layers: blockchain architecture, blockchain data structure, and blockchain storage engine, where block architecture indicates how to record transactions on a distributed ledger, blockchain data structure refers to the internal structure of each block, and blockchain storage engine specifies the storage form of data on the blockchain system. For each layer, the works advancing the state-of-the-art are discussed together with technical challenges. Furthermore, we lay out the future research directions for the blockchain data management systems.
\end{abstract}


%\ccsdesc[500]{Computer systems organization~Embedded systems}


\maketitle

\section{Introduction}
\label{sec:Introduction}

%Bingzhe's revision
Blockchain systems, such as Bitcoin~\cite{nakamoto2019bitcoin}, Ethereum~\cite{buterin2014next}, and HyperLedger~\cite{cachin2016architecture}, have received extensive attention in both academia and industries. A blockchain system can be regarded as a public ledger and transactions are committed to this ledger based on asymmetric cryptography and distributed consensus mechanisms. Blockchain technology generally has key characteristics of decentralization, persistency, anonymity, and auditability~\cite{zheng2017overview}.

%Bingzhe's revision
Currently, blockchain systems can be classified into three categories based on their chain structures: standard blockchain, hybrid blockchain, and DAG-based blockchain. Standard blockchain systems, such as Bitcoin and Ethereum, store all committed transactions in a chain list of blocks. The chain grows by appending new blocks continuously. In a standard blockchain system, any device can participate in the blockchain network as a node and maintains a full copy of the chain. For privacy protection, another architecture called hybrid blockchain, such as Hyperledger Fabric~\cite{cachin2016architecture}, and Ripple~\cite{armknecht2015ripple}, is proposed to maintain multiple chains in the network, and each node is only allowed to access the data in specified chains. Different chains cooperate together to exchange information and issue cross-chain transactions. 
Different from standard blockchain and hybrid blockchain, DAG-based blockchain systems, such as DagCoin~\cite{lerner2015dagcoin}, IOTA~\cite{popov2016tangle}, and Byteball~\cite{churyumov2016byteball}, utilize Directed-Acyclic-Graph (DAG) rather than block list as the basic data structure to manage transactions. In the DAG structure, each node is a transaction. When a transaction comes, it will connect two or more nodes in the DAG-based on node selection algorithms.

%Bingzhe's revision
Though blockchain systems have shown great potential in various areas, they suffer from two critical issues: security and scalability. Blockchain systems promise their security based on cryptography and consensus algorithms. However, many of these security mechanisms have theoretical weaknesses that expose them to various attacks, such as Malware attacks, Eclipse attacks, and Time jacking. Scalability is another issue that seriously limits the deployment of blockchain systems from several perspectives as follows. First, a blockchain system usually has a very low throughput, ranging from a few to dozens of transactions per second (TPS), which cannot deal with high-frequency trading scenarios. 
Second, with continuous operations of a blockchain system, the accumulative and excessive data load becomes a severe problem~\cite{timmurphy.org}, which not only slows down the transaction verification process, but also requires longer network broadcasting time, thus, further limiting system throughput. Moreover, to participate in a transaction verification process, a node needs to synchronize all data of a blockchain system. As a result, this excessive data load of large-scale blockchain systems may hinder the participation of many resource-limited nodes. 
%--verification, synchronous 
Third, the query process of a blockchain system is very time-consuming. 
%A highly efficient query engine for blockchain systems is urgently needed. 
To apply blockchain systems to applications in various fields~\cite{bonneau2015sok}, the query processing mechanism needs to be improved to provide low processing and response time with integrated query functions and high data security.

Some research studies investigate to improve the security and scalability of blockchain systems. For the security issue, many survey papers summarize the security problems and potential security threats in the current blockchain systems and introduced solutions to these security problems. For example, Lin et al.~\cite{lin2017survey} introduce several typical security issues in blockchain systems and summarize some of the current security challenges that need to be resolved. Li et al.~\cite{li2020survey} not only raise security issues and summarize solutions but also suggest some future directions in the security area. Anita et al.~\cite{8944615} summarize vulnerable attacks and different approaches against these attacks for the blockchain systems with a taxonomy of the attacks and a review of the countermeasures. Liu et al.~\cite{liu2019survey} focus on dealing with the security assurance of blockchain smart contracts and the correctness verification of blockchain smart contracts. 
The scalability issue of blockchain systems includes low throughput, excessive data load, and inefficient query engines. All these issues are highly related to the data management system of blockchain. However, to the best of our knowledge, no survey papers provide a thorough investigation of blockchain data management.

In this paper, we survey the technologies related to blockchain data management from three aspects, namely blockchain architecture, blockchain data structure, and blockchain storage engine. 
%The blockchain architecture is a description of how to record transactions on a distributed ledger. The blockchain data structure includes structures that are used to manage transactions on a chain. The blockchain storage engine refers to the underlying data storage engine of each blockchain node.
Specifically, in the blockchain architecture, our focus is on the techniques for how to record transactions on a distributed ledger; for the blockchain data structure, we summarize the structures that are used to manage transactions on a chain; in the blockchain storage engine,  our focus is on underlying data storage techniques for blockchain nodes. We introduce efforts that have been performed from these three aspects for three typical blockchain system designs: standard blockchain, hybrid blockchain, and DAG-based blockchain. Moreover, 
%we summarize some studies applying blockchains to different application scenarios. Finally, 
some potential research directions are discussed for blockchain data management.
%optimization directions are concluded for the data management system of blockchain.

%{\color{red}After summarizing the optimization methods for the data management system, we found that different methods target different blockchain systems, and the methods used are very different. A simple example is that some optimizations for Ethereum may be focus on smart contracts, but Bitcoin does not have the concept of smart contracts, so there is no way to intuitively compare the pros and cons of methods. In other words, there is no single method for all types of blockchain systems. this is a little strange here.} 

%(paper structure)
The rest of the paper is organized as follows. Section~\ref{sec:Background} introduces background information for standard blockchain, hybrid blockchain, and DAG-based blockchain.
Sections ~\ref{sec:standard}, ~\ref{sec:hybrid} and ~\ref{sec:dag} survey the data management techniques for standard blockchain, hybrid blockchain and DAG-based blockchain, respectively.
%Section~\ref{sec:standard} introduces the optimizations for standard blockchain system from three aspects. Section~\ref{sec:hybrid} introduces the optimizations for hybrid blockchain system and Section~\ref{sec:dag} introduces the optimizations for DAG-based blockchain system. 
Section~\ref{sec:app} summarizes the database techniques for blockchain. Section~\ref{sec:Related} describes the related work and Section~\ref{sec:Conclusion} concludes this paper.

\section{Background}
\label{sec:Background}

\subsection{Blockchain Data Management Stack}

\begin{figure}[ht]
  \centering
  \includegraphics[width=\linewidth]{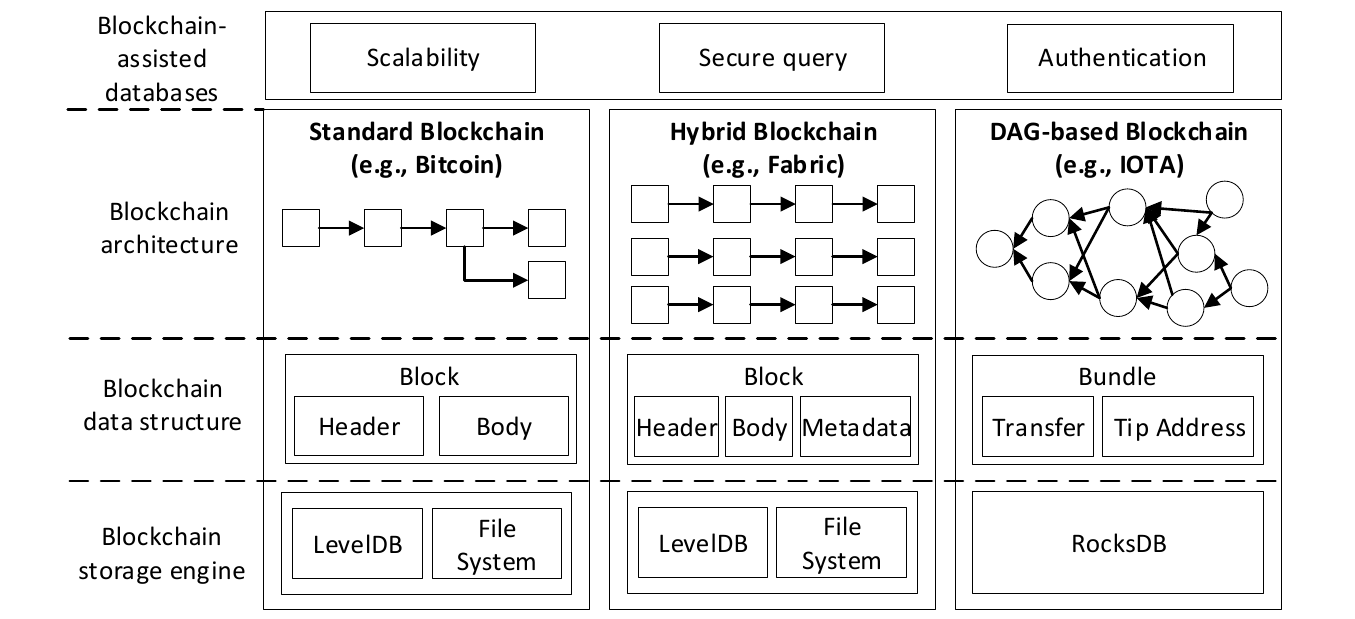}
  \caption{The layers of blockchain data management stack.}
  %\Description{A woman and a girl in white dresses sit in an open car.}
  \label{pic1}
\end{figure}

As shown in Figure~\ref{pic1}, the blockchain data management stack can be separated into four levels: the blockchain-assisted databases, blockchain architecture, blockchain data structure, and block storage engine. The blockchain architecture refers to the overall organizational structure of the blockchain network. Three typical blockchain designs (i.e. standard blockchain, hybrid blockchain, and DAG-based blockchain) use different architectures to maintain the distributed ledger. For instance, the standard blockchain adopts a list chain structure, the hybrid blockchain maintains several chains, and the DAG-based blockchain utilizes a DAG-based Tangle. The blockchain data structure refers to how the blockchain network organizes its transactions. The transactions of a blockchain system can either be organized into blocks by a mining process (Proof-of-Work (PoW) or Proof-of-Stake (PoS)) like in the standard and the hybrid blockchains or be directly attached to the network as a single node and verified by other nodes' confirmations like in the DAG-based blockchain. The blockchain storage engine is used to store and manage blockchain data on blockchain storage nodes. Blockchain systems usually adopt a database as the storage engine to maintain all data. Different blockchain systems may choose different database systems. Some blockchains choose a combination of a database and a file system to ensure data integrity~\cite{nakamoto2019bitcoin}.

Benefiting from the popular characteristics of the blockchain systems, people have deployed them as assistant database to various scenarios, such as information sharing and, private data protection. Depending on the different emphasis, we further divide the works which propose to adopt blockchain system to assist the design of databases from three aspects: scalability, secure query, and authentication.

During the investigation process, we start from the four perspectives of the data management system, namely blockchain architecture, blockchain data structure, blockchain storage engine, and blockchain-assisted databases. Based on these four aspects, according to the "standard blockchain", "hybrid blockchain", and "DAG-based blockchain" keyword search, we focus on the work done from the data management level and summarize the optimization work done for the data organization of blockchain systems.

\subsection{Standard Blockchain}

A standard blockchain system adopts a single chain list as its ledger. Transactions of the system are organized into a block and verified by a mining process. The verified blocks are added to the chain in chronological order. Blocks are linked according to their parent block hash in the block header until reaching the genesis block. 

A transaction block is the basic data structure in a standard blockchain system and consists of a block header and block body. Block metadata is stored in the block header, and the transactions are stored in the block body. The contents of the block header of different blockchain systems are slightly different. These contents can be roughly divided into two sets. One set includes mining-related information such as timestamp, difficulty, and nonce, and the other set includes Merkle root, parent block hash, and block version. 

The complexity of the data in the block header in standard blockchain systems is determined by its functional complexity. For example, the function of the Bitcoin system only includes the transfer of digital currencies, and thus the data in its blocks are relatively simple. Ethereum adds the concept of uncle blocks. Its block header also contains uncle hash. Therefore, the implementations of blockchain systems vary based on their different functions.

For the blockchain storage engine, a standard blockchain system usually adopts a file system to maintain its binary block data and a key-value database to maintain its metadata (e.g., Merkle tree and account information). For instance, Ethereum manages all its binary block data as normal files and adopts the key-value database LevelDB~\cite{ghemawat2011leveldb} to maintain its Merkle tree data to serve for account verification and query processes.

\subsection{Hybrid Blockchain}

Different with standard blockchain systems, hybrid blockchain systems are not public to all nodes but still offer blockchain features such as integrity, transparency, and security. The hybrid blockchain members can decide who can participate in the blockchain or which transactions are public. The blockchain architecture of a hybrid blockchain can have either one single listed chain as a standard blockchain system or several alliance chains. For the alliance chains, each chain corresponds to a set of ledgers, so each peer node in the blockchain may maintain multiple sets of ledgers. The structure of each chain is the same as a standard blockchain.

Similar to a standard blockchain system, a hybrid blockchain system also maintains transactions with the data structure of blocks. Each block mainly includes three parts: the relevant data of the block, the data related to its previous block, and the block metadata. The data of the block generally includes block hash and account number. Block metadata mainly includes the block creation time and signature.

For the blockchain storage engine, hybrid blockchain usually adopts a combination of databases and file systems. For instance, the storage engine of Hyperledger Fabric~\cite{cachin2016architecture} is composed of LevelDB and a file system. The storage engine of Ripple~\cite{armknecht2015ripple} is composed of a relational database (SQLite) and a key-value store.

\subsection{DAG-based Blockchain}
The blockchain architecture of a DAG-based blockchain system is based on Tangle~\cite{popov2016tangle}. Within Tangle, a DAG-based blockchain organizes all transactions in a DAG rather than packing transactions into blocks like in Bitcoin and Ethereum. Due to the non-block based structure, DAG-based blockchains are not limited by block generation. Therefore, DAG-based blockchains have great potential to provide higher throughput and lower latency than block-based blockchains.

For the data structure, in Tangle~\cite{attias2018tangle,popov2019equilibria}, each transaction is represented by a node. All the transactions are issued by clients through a selective proof-of-work (PoW) process. Leaf nodes (i.e., transactions that have not been approved by latter transactions) are called tips. For a transaction to be added, the system must select and approve two tips. During the tip selection and approval process, a random walking algorithm is used to choose tips for approval, and a transaction validation algorithm is performed to verify the transactions along the validation path to avoid double-spending.

For the blockchain storage engine, DAG-based blockchain usually adopts a relational or a key-value database to maintain its transactions and account information. For example, IOTA~\cite{popov2016tangle} adopts RocksDB to store all the transactions, the account information, and their corresponding addresses. Byteball~\cite{churyumov2016byteball} uses SQLite to store various types of data in different tables, such as the units table, to store the basic information of each unit.

%感觉这段要表达的意思并不是很明确，是不是可以这样总结，说下这三种blockchain的标志性特点之后再说下适合的应用场景？可能不太好找
%\subsection{Summary of the three blockchains}
%All blockchain systems follow the principle of decentralization and distinguish them through different blockchain architectures. 
In summary, Standard Blockchain is a kind of chained data structure that combines data blocks in chronological order. Each block composes multiple transactions and is generated by a mining process. Blocks are linked according to the parent block hash in the block header, all the way to the genesis block. Hybrid Blockchain supports multiple alliance chains, each chain corresponds to a set of ledgers. So each peer node in the blockchain may maintain multiple sets of ledgers, and the structure of each chain is the same as the standard blockchain. DAG-based Blockchain is based on the DAG structure. In the graph, each node is a transaction pointing to one direction and finally links to a root node. Standard Blockchain is widely used in the fields of digital currency, financial asset transaction settlement, deposit certificates, and anti-counterfeiting data services. Hybrid Blockchain is suitable for multi-organizational information exchange scenarios which can realize node transactions within and outside the organization. Since there are almost no transaction fees, DAG-based Blockchain is more suitable for the Internet of Things environment.

%Standard Blockchain广泛应用于数字货币、金融资产的交易结算存证防伪数据服务等领域，Hybrid Blockchain适用于多组织场景，可以实现组织内和组织外节点交易，DAG-based Blockchain由于其小额支付的特性比较适用于物联网环境。

%\textcolor{red}{ Another difference is that in the Standard Blockchain, multiple transactions are packaged in the same block, and then these blocks are sequentially linked to each other to form a chain. In Hybrid Blockchain, a chain is maintained within multiple alliances, each containing multiple blocks. Especially,, in the DAG-based Blockchain, only one transaction is stored in a block. }

%\input{3_Standard}
%\input{4_Hybrid}
%\input{5_DAG-based}
%\input{6_application}
%\input{7_Related}
%\input{8_Con}



\section{Standard Blockchain}
\label{sec:standard}

 In this section, we first survey data management techniques for standard blockchain systems in terms of blockchain architecture, blockchain data structure, and blockchain storage engine. Then we list several future optimization directions for data management in standard blockchain systems.

\subsection{Optimization for Standard Blockchain Architecture}

%fig2
\begin{figure}[h]
  \centering
  \includegraphics[width=\linewidth]{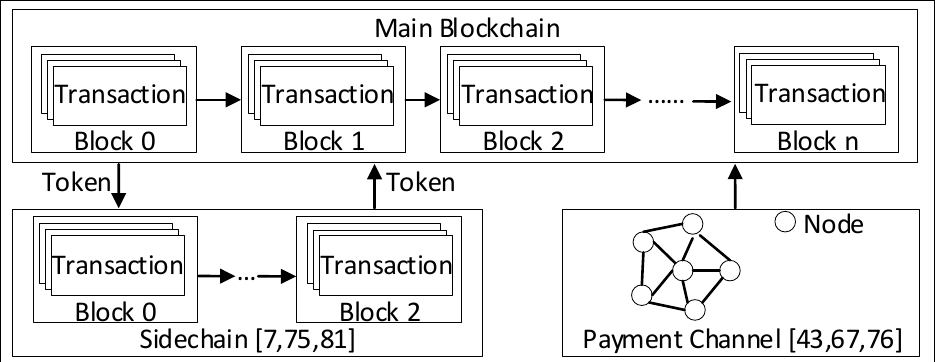}
  \caption{Optimization for the standard blockchain architecture.}
  %\Description{A woman and a girl in white dresses sit in an open car.}
  \label{pic2}
\end{figure}

%(table1)
\begin{table*}
   \caption{Overview of existing optimization papers for standard blockchain's architecture.}
   \label{tab:optimization for chain structure}
    \resizebox{\textwidth}{!}{
    \begin{tabular}{ccccccc}
\toprule
  Target problem                                                                                                               &  Solution                                                                                                                                  &  Authors                                                                          &  Year                                                                          &  Weakness \\\midrule 

                                                                                                                                          &                                                                                                                                                 &  Back et al.~\cite{back2014enabling}                              &  2014                                                                          &  Lack of transaction details   \\  

&                                                                                                                                                 &  Poon et al.~\cite{poon2017plasma}                               &  2017                                                                          &  Lack of transaction details   \\  

                                                                                                                                          &  \multirow{-3}{*}{ \begin{tabular}[c]{@{}l@{}}sidechain\end{tabular}}                                     &  Regnath et al.~\cite{DBLP:conf/iccad/RegnathS18}                              &  2018                                                                           &  Lack of transaction details     \\ \cline{2-5}

                                                                                                                                          &                                                                                                                                                 &  Poon et al.~\cite{poon2016bitcoin}                                &  2016                                                                         &  Insufficient security \\

&                                                                                                                                                 &   R. Network~\cite{network2018raiden}                               &  2016                                                                         &  Insufficient security \\

\multirow{-6}{*}{ \begin{tabular}[c]{@{}l@{}}The throughput bottleneck\end{tabular}}   &  \multirow{-3}{*}{ \begin{tabular}[c]{@{}l@{}}micropayment channel\end{tabular}}               &  Kim et al.~\cite{DBLP:conf/usenix/KimJJBS21}                           &  2021                                                                        &  Insufficient security    \\
\bottomrule
\end{tabular}  }         
\end{table*}

The previous studies that optimize the standard blockchain architecture mainly aim at solving the throughput bottleneck. The essential idea of these works is to reduce the burden on the main network, such as offloading some complex transactions into off-chain execution. To achieve this, the techniques can be classified into two categories: sidechain~\cite{back2014enabling,poon2017plasma} and micropayment channel~\cite{poon2016bitcoin,network2018raiden}. Table~\ref{tab:optimization for chain structure} and Figure~\ref{pic2} summarize these optimization works for the standard blockchain architecture. As shown in Figure~\ref{pic2}, the main idea of the sidechain is to build a small isolated chain with the same functionalities as the mainchain so that some transactions can first be carried out on the sidechain and then transferred to the mainchain. By doing so, some transactions are offloaded to the sidechains, thereby reducing the workload burden on the mainchain. 
In another category, separate micropayment channels are adopted to conduct off-chain transactions between various nodes, and only the final results are recorded on the mainchain. 

%----1214-14:07
%侧链是另一个完整的区块链，帮助主链来处理计算和任务
As a smaller isolated blockchain system, a sidechain is an independent chain with its own consensus mechanism and transaction type and runs its smart contract. Part of the transaction calculations and tasks of the mainchain is offloaded to the sidechain for processing.
The basis for the sidechain is an SPV-based (Simplified Payment Verification)~\cite{back2014enabling}  decentralized two-way peg technology. In the SPV mode, a user will first issue a special transaction to transfer certain of its own tokens (digital assets) to a dedicated address on the mainchain. Then, this address will issue an SPV transaction to the sidechain to make these tokens circulate on the sidechain. The left tokens of the user on the sidechain can be organized into another transaction and be returned to the mainchain by sending the transaction to the dedicated address again. 

Element chain~\cite{back2014enabling} is a sidechain scheme to offload the heavy data workload for the mainchain of Bitcoin. Element chain achieves a two-way anchoring between itself and the Bitcoin mainchain. It not only inherits the token transaction function of the mainchain but also supports other application scenarios by running smart contracts. Plasma~\cite{poon2017plasma} is a sidechain designed for the Ethereum mainchain. A sidechain root in Plasma is a smart contract running on the mainchain, which records the rules and the state hash of the sidechain. Multiple child chains can be generated from the root, which expands continuously and finally becomes a tree structure. Users can create a ledger on the Plasma chain and achieve asset-transfer between the Plasma chain and the Ethereum mainchain via the root.
Regnath et al.~\cite{DBLP:conf/iccad/RegnathS18} reduce verification steps by adding additional backlinks and enabled embedded devices to verify blockchain content using only a few kilobytes of RAM. The conventional block structure is extended to only connect a block to its direct predecessor, by a leap-hash allowing the user to traverse the blockchain with a reduced amount of steps to verify the inclusion and integrity of block data. By doing so, transaction complexity decreases, and throughput increases, but at the same time it brings additional resource consumption.

%临时的链下交易通道，将一些交易转移到该通道，以达到减少主链交易量的同时提高整个系统交易吞吐量的效果。
%payment channel 
Unlike a sidechain which can be regarded as the currency circulation between two independent blockchains, the micropayment channel is a transaction channel created between two nodes.  In a decentralized system, transactions are sent over a network of micropayment channels (a.k.a. payment channels or transaction channels) whose transfers of value occur off-blockchain.

Two typical micropayment channel implementations are Lightning Network~\cite{poon2016bitcoin} and Raiden Network~\cite{network2018raiden}.
The Lightning Network~\cite{poon2016bitcoin} adds another layer to Bitcoin’s blockchain, which enables users to create micropayment channels between any two parties. The bidirectional micropayment channel is contracts encumbered by hashlocks and timelocks. It is possible to clear transactions over a multi-hop payment network using a series of decrementing timelocks. Meanwhile, the Bitcoin transaction script in the channel is an implementation similar to "smart contract" to enable systems without trusted custodial clearinghouses or escrow services. These channels can exist as needed, and since they are established between two nodes, transactions are almost instant, the fees will be extremely low or even zero.

Raiden Network~\cite{network2018raiden}, similar to Lightning Network, is the micropayment channel of Ethereum and is designed to be used with any ERC20 compatible token. Raiden has four smart contracts, used to impose rule constraints on off-chain transactions. Currently, the routing in Raiden works in a simple manner. Unlike the Lightning Network that establishes a channel between two nodes, each node in the Raiden Network has a global view of the network. It knows the initial capacity of each channel by watching for the deposit events happening on the chain. Each node tries to forward the transfer through the shortest path with enough capacity to the target. If the transfer reaches the target and the protocol is followed properly then all the pending transfers in the path will be unlocked and the node balances will be updated.

Another micropayment channel implementation aims at executing transactions in the form of smart contracts off-chain. Kim et al.~\cite{DBLP:conf/usenix/KimJJBS21} recorded the historical Ethereum state that was used when a transaction was originally executed. The transaction can then be replayed in isolation for testing purposes. In addition, the recorder reorganizes the Ethereum world state into transaction-relevant substates to execute smart contracts in parallel.

%缺点
Although the above methods can offload transactions to sidechains or micropayment channels and improve the throughput of mainchain for a blockchain system, off-chain transactions may face security issues, such as node cannot confirm the accuracy of transactions. Because transactions are performed off-chain without the consensus verification process of the blockchain system, some malicious transactions and tampering under the data chain may happen. At the same time, since only the results of these transactions are sent to mainchain, the specific details of transactions are missed.
Once an error occurs in the account balance in the latest state of the blockchain, historical transaction information needs to be traced back. Some off-chain transactions will not be recorded, and some nodes may be malicious or have suffered security attacks. In this case, there would be impossible to determine how many malicious transactions occurred among the results of an on-chain transaction.

\subsection{Optimization for Standard Blockchain Data Structure}

\begin{figure}[!t]
% \centerings
\includegraphics[width=\linewidth,scale=1.00]{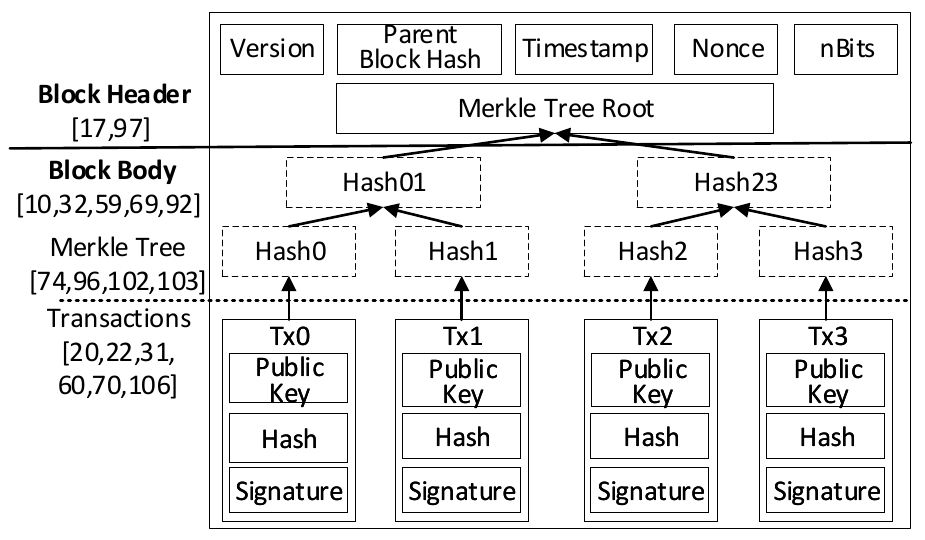}
\caption{Optimization for data structure of standard blockchain.}
\label{pic3}
\end{figure}

%table2
\begin{table*}
    \caption{A brief overview of existing optimization papers of standard blockchain's data structure.}
    \label{tab:optimization for data structure}
   \resizebox{\textwidth}{!}{
    \begin{tabular}{ccccccc}
\toprule
 Target problem                                                                                                                    &  Solution                                                                                                                         &  Authors                                                                                                   &  Year                                                                                 &  Weakness                                                                                                                                                               \\ \midrule
                                                                                                                                              &                                                                                                                                       &  Palai et al.~\cite{palai2018empowering}                                                  &  2018                                                                                 &   \begin{tabular}[c]{@{}l@{}}Only for transactions\\ involving transferable entities\end{tabular}                                \\  
                                                                                                                                              &                                                                                                                                       &  Gao et al.~\cite{gao2018blockchain}                                                       &  2019                                                                                 &   Cannot retain the historical traceability                                                                                                                  \\                                                                                         
                                                                                                                                              &                                                                                                                                       &  Bruce et al.~\cite{bruce2014mini}                                                            &  2014                                                                                 &  Insufficient security                                                                                                                                                \\ 
                                                                                                                                              &                                                                                                                                       &  Bitcoinwiki.~\cite{wiki2015scalability}                                                        &  2019                                                                                 &  Data loss                                                                                                                                          \\  
                                                                                                                                              &                                                                                                                                       &  Cross et al.~\cite{lombrozo2015segregated}                                            &  2015                                                                                 &  Additional storage load                                                                                                                                  \\ 
                                                                                                                                              &   \multirow{-7}{*}{ \begin{tabular}[c]{@{}l@{}}Partial storage\end{tabular}}                 &  Dai et al.~\cite{dai2019jidar}                                                                    &  2019                                                                                 &  Increases the cost                                                                                                                                        \\ \cline{2-5}

                                                                                                                                              &                                                                                                                                        & Xu et al.~\cite{xu2018section}                                                                  &  2018                                                                                 &  Data loss                                                                                                                                         \\ 
                                                                                                                                              &                                                                                                                                        &  Bip152.~\cite{corallo2016bip152}                                                              &  2019                                                                               &  Increases the cost                                                                                                                                           \\ 
                                                                                                                                              &                                                                                                                                        &  Ding et al.~\cite{ding2019txilm}                                                                &  2019                                                                               &  Data loss                                                                                                                                          \\ 
                                                                                                                                              &                                                                                                                                         &  Patsonakis et al.~\cite{patsonakis2019alternative}                                    &  2019                                                                                &  Total cost increase                                                                                                                                                  \\ 
                                                                                                                                              &                                                                                                                                         &  Liu et al.~\cite{DBLP:journals/tii/LiuWLX19}                                    &  2019                                                                                &     Data loss                                                                                                                                               \\ 
                                                                                                                                              &                                                                                                                                         &  Ge et al.~\cite{DBLP:journals/jsa/GeML20}                                    &  2020                                                                                &     Data loss                                                                                                                                               \\ 

                                                                                                                                              &   \multirow{-7}{*}{ \begin{tabular}[c]{@{}l@{}}Block compression\end{tabular}}         &  Zheng et al.~\cite{zheng2018innovative}                                                     &  2018                                                                                 &  Increases the cost                                                                                                                                                \\ \cline{2-5}

  \multirow{-14}{*}{ \begin{tabular}[c]{@{}l@{}}Excessive data load\end{tabular}}               &   \multirow{-1}{*}{ \begin{tabular}[c]{@{}l@{}}Simplified ADS\end{tabular}}                                                                                                            &  Ponnapalli et al.~\cite{ponnapalli2019scalable}                                          &  2019                                                                                &  Decrease in decentralization                                                                                                                          \\     \cline{1-5}

&                                                                                                                                         &  Zhang et al.~\cite{DBLP:conf/icde/ZhangXWXC21}                                    &  2021                                                                                &     Data loss                                                                                                                                               \\

                                                                                                                                              &                                                                                                                                        &  Zhang et al. ~\cite{zhang2019gem}                                                            &  2019                                                                              &  High ADS maintenance costs                                                                                                                                           \\ 
 \multirow{-3}{*}{ \begin{tabular}[c]{@{}l@{}}Inefficient query performance\end{tabular}}  &  \multirow{-3}{*}{ \begin{tabular}[c]{@{}l@{}}Redesigned ADS\end{tabular}}                &  Xu et al.~\cite{xu2019vchain}                                                                     &  2019                                                                               &  Increases the cost                                                                                                                                              \\ \bottomrule
\end{tabular} }          
\end{table*}

Table~\ref{tab:optimization for data structure} and Figure~\ref{pic3} provide a summary of studies that optimize standard blockchain systems by redesigning their on-disk and in-memory data structures. Figure~\ref{pic3} shows the internal structure of a block, which mainly consists of a block header and block body. In this figure, we classify these studies into optimization for the block header~\cite{xu2018section, corallo2016bip152} and optimization for the block body~\cite{palai2018empowering, bruce2014mini, wiki2015scalability}. For the optimization of the block body, most of the works~\cite{gao2018blockchain, lombrozo2015segregated, dai2019jidar, ding2019txilm, patsonakis2019alternative} focus on improving the transaction part, and some~\cite{ponnapalli2019scalable, zhang2019gem, xu2019vchain} focus on improving the structure of the Merkle Tree.

\subsubsection{Optimization for Excessive Data Load}

To mitigate the excessive data load of standard blockchain systems, there are mainly three approaches, namely, partial storage, data compression, and simplified authenticated data structure (ADS). The main idea of partial storage is to reduce the block size by removing a certain part of a block. In general, the removed part is the block body since some simple verification is done only depending on the block header, and most users are not interested in the content of transactions issued by other users. For block compression, methods such as changing the smart contract form, encoding data, etc., are proposed to compress the data in blocks. By doing so, the block size can be significantly reduced. ADSs such as Merkle trees are widely used in blockchain systems to verify if the data is correct and up-to-date. ADSs incur significant storage overhead due to random I/O, serialization, and hashing operations. Several studies propose to shard the tree and distribute the shards to  different blockchain nodes~\cite{ponnapalli2019scalable}.

%块汇总，只存储一系列块的更改数据
$\bullet${\it Partial storage.}
For partial storage, Palai et al.~\cite{palai2018empowering} propose to perform the block summary of outdated blocks and only broadcast the block summary in blockchain systems. The block summary can store all the input resources for the given blocks and the total change introduced by these blocks. The proposed method allows resource-restricted light nodes to store an integrated blockchain so that it can validate transactions independently, which ultimately reduces dependency on full nodes.
Though the work of verifying transactions only needs to include part of historical transactions, once there is a request to obtain information about other transactions, the node has to seek the assistance of other full nodes.

%定期汇总块，总结历史交易数据
Gao et al.~\cite{gao2018blockchain} propose to offload the block summary to an InterPlanetary File System (IPFS) and only maintain the updated transactions in the current blockchain system. The scheme mainly achieves space optimization by summarizing blocks regularly, excluding the expired Transaction Output (TXO) that has no effect on verification. The Unspent Transaction Output (UTXO) will be saved as a file on the IPFS network during optimization. In this way, the miner node can quickly get the file and start mining. The scheme summarizes the historical transaction data regularly, and eliminates the Spent Transaction Output (STXO) to effectively reduce the demand for node storage space.

%删除旧的交易，只保留区块头,但也保存最新区块的交易数据
Bruce et al.~\cite{bruce2014mini} propose a P2P crypto-currency scheme where old transactions can be removed from the network, thereby reducing the amount of storage. Since nodes only require the latest portion of the blockchain to sync with the network, they call this portion of the chain the “mini-blockchain”. Because the old transaction records are deleted, the account tree is used to fulfill the task of managing account balances and recording the ownership of coins. This also brings some potential benefits, such as faster transaction speed.
On the other hand, deleting transaction records also brings potential security problems. Once some malicious transaction records are deleted, the account balance cannot be called back and corrected.

%只保留区块头

Simplified payment verification (SPV)~\cite{wiki2015scalability} is proposed to only store the elementary information of each block, including hash, proof of work, etc. No matter how large the future transaction volume is, the size of the block header always remains the same (80 bytes), which significantly saves storage space and reduces the burden on back-end users. Therefore, a user with limited-resource equipment is capable of loading a blockchain system under normal circumstances.

Segregated witness~\cite{lombrozo2015segregated} is to remove the signature data in a transaction to reduce its size so that the block can accommodate more transactions, thereby reducing the overall data load. A new data structure called "witness" is defined to contain data required to check transaction validity but without data that determines transaction effects. In particular, scripts and signatures are moved into this new structure. At the same time, the witness also brings additional management and storage overhead.

Jigsaw-like Data Reduction (Jidar)~\cite{dai2019jidar} is a data reduction approach for the Bitcoin system. The main idea of Jidar is to allow nodes only to store transactions and relevant Merkle branches that they are interested in. Thus, it can alleviate the storage pressure of each node. The related Merkle branches are provided with a transaction to verify the validity of the newly proposed transaction. Given the requirement that a user wants to put all the fragments stored in different nodes into a complete block, just like stitching all the pieces into a complete jigsaw-puzzle picture, a mechanism of querying full data is added to Jidar. This design neither has any trust assumptions for the Bitcoin system, nor sacrifices the original properties of the blockchain.
 
$\bullet${\it Block compression.}
Block compression reduces the data load by redesigning the block structure. 
Xu et al.~\cite{xu2018section} present section-blockchain, a new blockchain protocol, to solve the oversize storage problem without compromising the security of a blockchain. There is no full node and lightweight node. All nodes equally contribute to the section-blockchain network. Block storage mechanism is redesigned based on the original Nakamoto blockchain, which allows the nodes only keeping (1) the block headers in the mainchain; (2) a certain number of blockchain fragments; (3) a certain number of Database Snapshot (DS); (4) a Fragment and Database snapshot Grabbing Routing Table (FDGRT). Experiments demonstrate that section-blockchain is efficient, remarkably reduces the storage footprint, and withstands sudden data loss from massive nodes.

%To solve the problem of composition 
In the scenario where blockchain nodes are small-memory, resource-constrained devices, Ge et al.~\cite{DBLP:journals/jsa/GeML20} redesign the blockchain constitution to mitigate the pressure of storage and calculation imposed on each node.
Their solution relies on a voting mechanism combining with a reputation evaluation scheme. The item of reputation tree is added into the block and the root of tree is recorded by block header to further increase throughput, but the safety is relatively weaker than previous designs.

Liu et al.~\cite{DBLP:journals/tii/LiuWLX19} present the concept of lightweight blockchain to solve the problem of excessive data load through two approaches. The first one is called Unrelated Block Offloading Filter (UBOF). Through the analysis of Unspent Transaction Output (UTXO), they propose the definition of Unrelated Blocks (UB). UBOF can detect and offload UBs, contributing to reducing the storage resources occupied by blockchain. The other is a lightweight data structure named LightBlock (LB). They propose to broadcast LB instead of the entire block after one block is generated.

Compact block relay, or BIP152~\cite{corallo2016bip152} (Bitcoin Improvement Proposal), reduces the amount of network requirement from P2P network nodes to broadcast blocks. The compact block relay changes the data structure of the original block in Bitcoin. As a result, a compact block only contains the header of the block and some short transaction IDs (TXIDs), which are used to match the transactions already available to the receiver. The receiver node will use the received information and the transactions in its memory pool to reconstruct the entire block instead of transmitting the entire block. Txilm~\cite{ding2019txilm} is a protocol based on BIP152 that compresses transactions in each block to save the bandwidth of the network.  In this protocol, a block carries the short hashes of TXIDs instead of completing transactions. Thus, the protocol achieves a higher block compression ratio. However, hash collisions more likely occur when a short hash is used. Therefore, Txilm optimizes the protocol using sorted transactions based on TXIDs to reduce the probability of hash collisions. Combined with the sorted transactions based on TXIDs, Txilm realizes 80 times of data size reduction compared with the original blockchains.

Patsonakis et al.~\cite{patsonakis2019alternative} introduce an alternative paradigm for developing smart contracts, in which the storage consumption is of constant size. This will significantly reduce the amount of storage consumed by smart contracts but cause a certain incentive imbalance due to the high cost. To solve this problem, they introduce recurring fees that are proportional to the state of smart contracts and adjustable by miners that maintain the network.  
At the same time, it also brings an increase in the total cost of the entire blockchain network.

%数据放入IPFS中，并返回IPFS哈希
Zheng et al.~\cite{zheng2018innovative} propose an IPFS-based blockchain data storage model. With this model, miners deposit transaction data into the IPFS network and pack the returned IPFS hash of the transaction into a block. Utilizing the characteristics of the IPFS network and the features of the IPFS hash, the blockchain data can be significantly reduced.

$\bullet${\it Simplified ADS.}
Different from the methods mentioned earlier, Ponnapalli et al.~\cite{ponnapalli2019scalable} present a new authenticated data structure (ADS) to reduce the excessive data load. A Distributed Merkle Patrica Tree (DMPT) is designed to reduce the storage and computation overheads of Merkle trees and their variants. The DMPT eliminates storage overhead by storing the entire tree in memory. A DMPT vertically shards the Merkle Partica tree across the memory of multiple nodes, stored in the memory of a different node, eliminating the I/O bottleneck. Since both reads and writes in Merkle trees operate on a vertical path down the tree, vertical sharding enables process reads and writes on different nodes in parallel. DMPTs further help reduce network bandwidth utilization by combining novel techniques such as witness compaction and node bagging. 
Although sharding MPT on different nodes can greatly increase the degree of parallelism, it reduces the degree of decentralization since a single node in the network cannot independently complete verification and query work.

In summary, although the methods of partial storage and block compression significantly reduce the data load and increase system throughput, they will lead to a certain amount of information loss. For example, due to the imperfection of the data information, the query process becomes difficult. The node can only be used with the assistance of full nodes to complete the query on the specific transaction details, otherwise, it can only verify whether the transaction is confirmed.

\subsubsection{Optimization for Query Engine}

Current blockchain systems lack a complete and efficient query system, the query efficiency is low and the provided query interfaces are limited. The underlying data storage system of most standard blockchain systems uses Log-structured Merge Tree (LSM-tree) based key-value stores, which sacrifice read performance in exchange for write performance. Moreover, the unstructured data storage system based on the Key-Value model does not support complex queries, which has become the main bottleneck to restrict query functions.

Xu et al.~\cite{xu2019vchain} propose a novel framework, called vChain, to alleviate the storage and computation cost of blockchain systems. The scheme employs verifiable queries to guarantee data integrity. To support verifiable Boolean range queries, they propose an accumulator-based authenticated data structure that enables dynamic aggregation over arbitrary query attributes. Two new types of indexes are developed to aggregate intra-block and inter-block data records for efficient query verification. They also propose an inverted prefix tree to accelerate the processing of a large number of subscription queries simultaneously. 
Although verifiable query ensures data integrity, it also brings high ADS maintenance costs.

By analyzing the performance of the existing techniques, Zhang et al.~\cite{zhang2019gem} propose a novel ADS, called $GEM^{2}-tree$, which is not only gas-efficient but also effective in supporting authenticated queries. To further reduce the ADS maintenance cost without sacrificing query performance, they also propose an optimized structure, $GEM^{2*}-tree$, by designing a two-level index structure.

Zhang et al.~\cite{DBLP:conf/icde/ZhangXWXC21} apply an authenticated query process, in which both the smart contract and the off-chain SP maintain an authenticated data structure (ADS) named Merkle inverted index. The Merkle inverted index is an inverted index in which each keyword corresponds to a Merkle B-tree (MBtree) that indexes the corresponding object IDs. The smart contract maintains only the root digest of each MB-tree in the Merkle inverted index. This idea comes with the observation that for the on-chain ADS, only the root digests are used during the authenticated keyword search, which reduced the storage burden.

In summary, although these techniques improve the system query performance, they may induce a large overhead on the storage side due to storing the entire ledger. Hence, it becomes critical to handle excessive data load in blockchain systems. Particularly, even though these methods can improve the query efficiency, due to a large amount of data, the throughput of blockchain systems still remains low.

\subsection{Optimization for Standard Blockchain Storage Engine}

\begin{figure}[!t]
% \centerings
\includegraphics[width=\linewidth,scale=1.00]{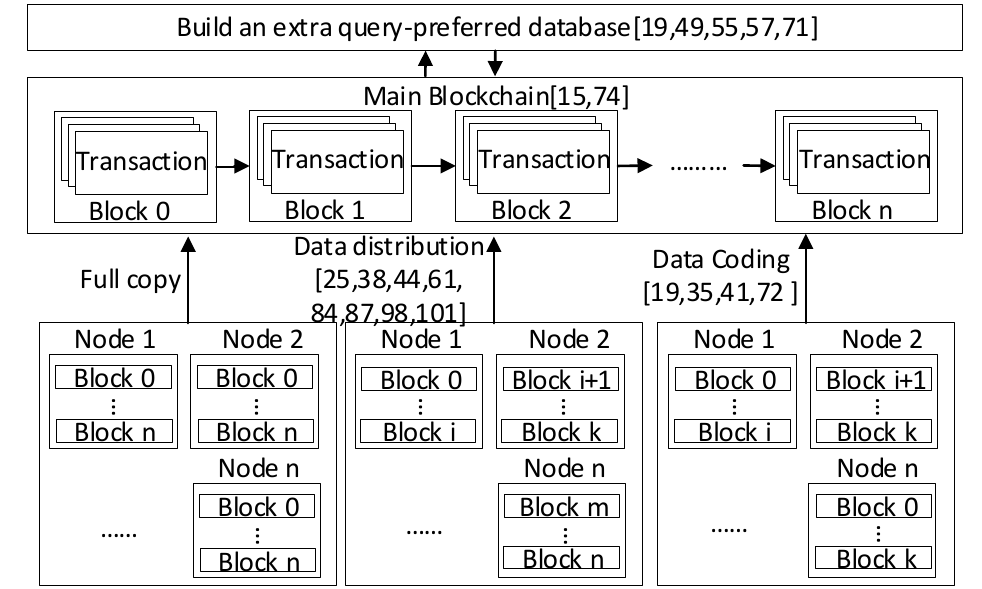}
\caption{Optimization for storage engine of standard blockchain.}
\label{pic4}
\end{figure}

%table3
 \begin{table*}  
    \caption{A brief overview of existing optimization papers of standard blockchain's storage engine.}
    \label{tab:optimization for storage engine}
     \resizebox{\textwidth}{!}{
    \begin{tabular}{ccccccc}
\toprule
  Target problem                                                                                                                   &  Solution                                                                                                                     &  Authors                                                                     &  Year                                                            &  Weakness                                                                                                    \\ \midrule

                                                                                                                                              &                                                                                                                                     &  Luu et al.~\cite{luu2016secure}                                &  2019                                                             &  Decrease in decentralization                                                                                       \\

                                                                                                                                              &                                                                                                                                     &  Kokoris et al.~\cite{kokoris2018omniledger}             &  2019                                                            &  Decrease in decentralization                                                                                       \\

                                                                                                                                              &                                                                                                                                     &  Zamani et al.~\cite{zamani2018rapidchain}             &  2019                                                           &  Decrease in decentralization                                                                                       \\

                                                                                                                                              &                                                                                                                                     &  Wang et al.~\cite{wang2019monoxide}                    &  2019                                                           &  Decrease in decentralization                                                                                       \\

                                                                                                                                              &                                                                                                                                     &  Han et al.~\cite{DBLP:conf/icde/HanXC18}                    &  2018                                                          &  Decrease in decentralization                                                                                       \\ 

                                                                                                                                              &                                                                                                                                     &  Tao et al.~\cite{DBLP:conf/icde/TaoLJNWL20}                    &  2020                                                          &  Increase in delay                                                                                      \\

                                                                                                                                              &                                                                                                                                     &  Drakatos et al.~\cite{DBLP:conf/icde/DrakatosDKKZ21}                    &  2021                                                          &  Increase the complexity                                                                                      \\ 

                                                                                                                                              &   \multirow{-8}{*}{ \begin{tabular}[c]{@{}l@{}}Data  distribution\end{tabular}}                 &  Xu et al.~\cite{xu2018cub}                                       &  2018                                                            &  Decrease in decentralization                                                                                       \\ \cline{2-5}

                                                                                                                                              &                                                                                                                                     &  Dai et al.~\cite{dai2018low}                                      &  2018                                                           &  Increase query complexity                                                                                       \\

                                                                                                                                             &                                                                                                                                      &  Guo et al.~\cite{guo2019design}                              &  2019                                                            &  Increase query complexity                                                                                       \\ 

                                                                                                                                             &                                                                                                                                      &  Jiang et al.~\cite{DBLP:journals/jsa/JiangLGYYSL20}                              &  2020                                                            &  Increase query complexity                                                                                       \\

\multirow{-8}{*}{ \begin{tabular}[c]{@{}l@{}}Excessive data load\end{tabular}}                   &  \multirow{-4}{*}{ \begin{tabular}[c]{@{}l@{}}Data coding\end{tabular}}                      &  Perard et al.~\cite{perard2018erasure}                     &  2018                                                            &  Increase query complexity                                                                                  \\ \cline{1-5}

                                                                                                                                             &                                                                                                                                     & Lesavich et al.~\cite{lesavich2017method}                 &  2017                                                           &  Too much process cost                                                                                  \\

                                                                                                                                             &                                                                                                                                     &  Li et al.~\cite{li2017etherql}                                     &  2017                                                            &  Too much process cost                                                                                       \\

                                                                                                                                             &                                                                                                                                     & Peng et al.~\cite{peng2019vql}                                  &  2019                                                           &  Too much process cost                                                             \\

                                                                                                                                             &                                                                                                                                     & Linoy et al.~\cite{DBLP:conf/icde/LinoyRS20}   &  2020                                                           &  Too much process cost                                                             \\  

\multirow{-5}{*}{ \begin{tabular}[c]{@{}l@{}}Inefficient query mechanism\end{tabular}}    &  \multirow{-5}{*}{ \begin{tabular}[c]{@{}l@{}}Connect  outside database\end{tabular}}    &  Hanley et al.~\cite{hanley2018managing}                  &  2018                                                          &  Excessive storage resource consumption                                                                                       \\ \cline{1-5}

                                                                                                                                             &                                                                                                                                      &  Ponnapalli et al.~\cite{DBLP:conf/usenix/PonnapalliSBMTC21}                              &  2021                                                            &  Increasing the complexity                                                                                    \\

\multirow{-2}{*}{ \begin{tabular}[c]{@{}l@{}}Low throughput\end{tabular}}                   &  \multirow{-2}{*}{ \begin{tabular}[c]{@{}l@{}}Reduce I/O of tx\end{tabular}}                      &  Chen et al.~\cite{chenzehao}                     &  2021                                                            &  Excessive storage resource consumption                                                                                   \\

\bottomrule
\end{tabular} }          
\end{table*}

Table~\ref{tab:optimization for storage engine} and Figure~\ref{pic4} provide the summary of works that optimize blockchain data load, throughput and query processing from the perspective of the blockchain storage engine. As shown in Figure~\ref{pic4},  the techniques for handling excessive data load are mainly categorized into  distributed storage and data coding. To provide an efficient query engine, several studies propose to build an extra query-preferred database.

\subsubsection{ Optimization for Excessive Data Load} 

$\bullet$ {\it Data Distribution.}  
The main idea of data sharding is that multiple nodes form a group with high trust to jointly store all data of a complete node. Most of the works use the idea of sharding~\cite{dang2019towards} to distribute the transaction processing overhead among multiple smaller groups of nodes.
The current sharding works are mainly divided into network and transaction sharding and state sharding. Through network and transaction sharding, blockchain nodes are divided into different shards, and each shard forms an independent process and gets consensus on different transaction subsets. In this way, a subset of transactions that are not connected to each other can be processed in parallel, and transaction throughput can be significantly improved by orders of magnitude. The idea of state sharding is to separate the entire storage area and make different shards to store different parts. Therefore, each node is only responsible for hosting its own shard data, rather than storing the complete blockchain states.

For network and transaction sharding, Elastico~\cite{luu2016secure} is the first sharding protocol for the permission-less blockchain. Technically, Elastico uniformly partitions or parallelizes the mining network into smaller committees, each of which processes a disjoint set of transactions (i.e., “shards”). Each committee has a reasonably small number of members so they can run a classical byzantine consensus protocol to decide their agreed set of transactions in parallel.
After that, OmniLedger~\cite{kokoris2018omniledger}, a more recent distributed ledger based on the Sharding technique, is proposed to solve the problems of Elastico. Specifically, to commit transactions atomically across shards, OmniLedger introduces an atomic commit protocol called Atomix. OmniLedger also adopts state blocks that are deployed along OmniLedger to minimize storage and update overhead.
Rapidchain~\cite{zamani2018rapidchain} reduces the amount of data exchange per transaction and does not need to gossip transactions to the entire network because of the usage of a fast cross-shard verification technique. Monoxide~\cite{wang2019monoxide} is a scale-out blockchain that proposes Asynchronous Consensus Zones and scales the blockchain linearly while maintaining decentralization and security.

Tao et al.~\cite{DBLP:conf/icde/TaoLJNWL20} implement a new distributed and dynamic sharding system. They mainly discuss a problem in sharding, which is that a small shard with few transactions tends to generate a large number of empty blocks resulting in a waste of mining power, while a large shard adversely affects parallel confirmations. An inter-shard merging algorithm with incentives to encourage small shards is proposed to merge and form a larger shard. On the other hand, different from the non-sharding systems, miners must communicate with each other to exchange their individual validation results to validate the transaction jointly.

%共识单元CU，不同节点一起维护一整个账本
As for state sharding, Xu et al.~\cite{xu2018cub} introduce a novel concept called Consensus Unit (CU), which organizes different nodes into one unit. The nodes in the same CU will work together and maintain at least an entire copy of blockchain data. Based on this idea, they further define a Blocks Assignment Optimization (BAO) problem to solve the optimal assignment of blocks such that the storage space is fully used and the query cost is minimized.
Han et al.~\cite{DBLP:conf/icde/HanXC18} develop a blockchain platform to support mobile devices. A novel concept called consensus unit which groups several single nodes in the network into a trusted cluster is introduced. The CU management module is used to monitor the status of the CU. The user can get the information of active peers and the ledger status of the current CU. When a user finds the ledger occupying too much storage space, it can raise optimizing storage requests with several sharding plans to CU peers.

Drakatos et al.~\cite{DBLP:conf/icde/DrakatosDKKZ21} propose Triabase which divide the chain nodes into two types: nodes that store the entire shared database, and the others that use the database for their own operations, such as sending query and update requests to the blockchain shared ledger. Triabase carries out machine learning at the edge, abstracts machine learning in primitive blocks that are subsequently stored and retrieved from the blockchain.

Though the data distribution method efficiently mitigates the storage pressure, it weakens the decentralization of blockchain systems. The data sharding method requires a certain trust relationship among nodes, which is not required by the original blockchain design to ensure that data is not tampered with.

%使用擦除码重建1 2 3
$\bullet$ {\it Data Coding.} 
Several studies propose to use erasure coding to process blockchain data and then perform distributed storage to mitigate the data load problem of a single node.
Perard et al.~\cite{perard2018erasure} introduce a new type of node, an erasure-code based low storage node, which stores coded fragments for each block of a blockchain. Similar to~\cite{perard2018erasure}, Dai et al.~\cite{dai2018low} adopt a network coded (NC) distributed storage (DS) framework to store a blockchain to reduce storage requirements. NC applied to DS can easily rebuild the whole ledger from a small number of such nodes compared with the method of simple replications. Correspondingly, when the system fetches these encoded blocks, it needs to perform decoding before the normal blockchain procedure.

Different from~\cite{perard2018erasure} and~\cite{dai2018low}, Guo et al.~\cite{guo2019design} propose a new storage mechanism based on a redundant residual number system to reduce the storage overhead of account information on each node instead of the whole ledger. In the proposed scheme, the information for each account is partially stored on each node as its residual number (RNS) related to a module assigned to the node. Since the modules can be an order of magnitude smaller than the normal account information, the storage utilization of the account information on each node can be dramatically reduced.

In UTXO-based blockchain, the increasing size of UTXO set will soon become the biggest obstacle for nodes to validate a high amount of transactions in time, and will eventually become the bottleneck to implement high-throughput blockchains. Jiang et al.~\cite{DBLP:journals/jsa/JiangLGYYSL20} try to minimize the footprint of unspent transactions with a shorter representation form of key and value. For the key, a shorter transaction ID and index is provided by intercepting a part of the original full key, but at the same time, duplicate keys may appear. Once there is a conflict problem, it will query an extra table, which stores the original full key and the corresponding value of these key-value pairs. In addition, a user usually has many UTXOs and the same user information is duplicated in each entry. For the value, a two-level hash mapping technique is proposed for eliminating the redundant data. Specifically, every unique user address is stored as a key in the second table whose value domain is set to null, and the value domain in the first table stores a pointer pointing to use address data. To adapt to the proposed method, some modifications have been made to the underlying database. Compared to the original, the read operation adds steps to access the extra table and the second hash table while the write operation adds steps to compare with possibly conflicting keys, write to the extra table, and access to the second hash table, which adds the complexity of the database.

%{\color{blue}(----the following sentence is very strange, I did not understand its purpose ---zhaoyan)---What I want to express is that for the proposed blockchain data organization method, several steps have been added to the reading and writing of the database.}

The methods of data coding also face an extremely large overhead to recover the entire ledger via some assistance since each node only stores a part of the whole ledger. Therefore, when a single point or range of data is inquired, the steps of restoring the original data are added, which increases the complexity of the system.

%{\color{green}{(the previous sentences are right? -Bingzhe)----revised}}
%For example, when a node needs to query data, it must first request for the assistance of other nodes to rebuild the entire ledger, resulting in lots of time and resources.
%The methods of data coding also face the problem of data information loss. When a node needs to query data, it must first request for the assistance of other nodes to rebuild the entire ledger, which will consume lots of time and resources. 

\subsubsection{Optimization for Query Engine}

To improve query efficiency, several techniques are proposed to build an additional external query-preferred database on blockchain systems. The design of these works mainly includes three parts: data monitoring and analyzing module, external database engine, and querying API. The data monitoring and analyzing module monitors the data of a blockchain system and exports the required data to an external query-preferred database engine. Then, a set of API is provided to serve the users' requests. In the following, we will introduce these implementations in detail.

%使用Galois Fields存储和检索
Lesavich et al.~\cite{lesavich2017method} propose a method for storage and a system for retrieval of blockchains with Galois Fields, respectively. One or more blocks for a blockchain are securely stored and retrieved with a modified Galois Fields on a cloud or peer-to-peer (P2P) communications network. The modified Galois Field provides additional layers for security and privacy for blockchains. The blocks are securely stored and retrieved for cryptocurrency transactions including, but not limited to, BITCOIN transactions and other cryptocurrency transactions.

%设置查询层（自动同步块数据到专用数据库中）
Li et al.~\cite{li2017etherql} develop EtherQL, an efficient query layer for Ethereum. EtherQL automatically synchronizes new block data in a timely manner and stores them in a dedicated database to ensure efficient and accurate queries. EtherQL provides highly efficient query primitives for analyzing blockchain data, including range queries and top-k queries, which can be flexibly integrated with other applications.

%提取底层交易，并且在数据库中重组
To provide efficient and verifiable data query services for blockchain-based systems, Peng et al.~\cite{peng2019vql} propose a Verifiable Query Layer (VQL) middleware. The middleware extracts transactions stored in the underlying blockchain system and reorganizes them in databases to provide various query services for public users. To prevent falsified data from being stored in the middleware, a cryptographic hash value is calculated for each constructed database. The database fingerprint including the hash value and some database properties will be first verified by miners and then stored in the blockchain.

Hanley et al.~\cite{hanley2018managing} present a blockchain system that uses offchain centralized data storage. The system provides instant access to patients' and medical professionals' medical records from anywhere. By assigning each medical record a pseudo-anonymous identifier, a second layer "blockchain" for each user can be created to allow for the rapid collection and data queries. The offchain pseudo-anonymous data storage allows data to remain unencrypted. Thus, it enables the rapid generation of anonymous medical datasets that can be used for machine learning and data mining on the data, potentially bringing many benefits to the healthcare industry.
As the data of the second-tier blockchain system will accumulate all the time, compared with other dedicated databases, the consumption of storage resources is too large and will become more serious with the increment of time.

To analysis a specific historic transaction, all preceding transactions in all preceding blocks need to be run, which includes the execution of smart contracts. Linoy et al.~\cite{DBLP:conf/icde/LinoyRS20} propose an Eidetic blockchain system that supports provenance for both data (blockchain states) and control flow (contracts). Eidetic captures the provenance of the contracts’ execution flow (execution flow provenance), their parameters, and the relevant blockchain states before and after each contract call (Why-provenance). This provenance information can then be used to query the execution flow across time in different granularity levels.

In summary, with the help of the query-preferred external databases, these designs make the query process of the blockchain system much efficient. However, the additional extra database induces a much higher complexity to the implementation of blockchain systems. Moreover, since the storage size of a blockchain system can be extremely large, exporting and reorganizing it with an additional external database takes much longer time and consumes many resources. As a result, these methods may further increase the data load and decrease the overall performance of blockchain systems.

\subsubsection{ Optimization for Low Throughput} Public blockchains suffer from low transaction throughput. Two popular public blockchains, Bitcoin and Ethereum, deal with only tens of transactions per second, limiting their deployment~\cite{eth}. To solve this problem, some research works are proposed to improve the execution speed of transactions by reducing the number of disk I/Os.

Processing a transaction involves executing the transaction and verifying the validation of the execution result. In Ethereum, both execution and verification require reading and writing system states that are stored in a Merkle tree on local storage. Ponnapalli et al.~\cite{DBLP:conf/usenix/PonnapalliSBMTC21} design a custom storage solution namely Distributed, Sharded Merkle Tree (DSM-TREE). DSM-TREE is designed in RAINBLOCK which deconstructs miners into three entities: storage nodes that store data, miners that process transactions, and I/O-Helpers that fetch data and proofs from storage nodes and provide them to miners. By having I/O-Helpers prefetch data on behalf of miners, I/O is removed from the critical path of transaction processing; moreover, multiple I/O-Helpers can prefetch data at the same time, which increases the I/O pluralism of the system.

As the blockchain network scales up, the systems face more severe storage performance degradation since inserting and accessing data from the storage layer become more expensive as the total size of the data increases. Chen et al.~\cite{chenzehao} propose BLock-LSM which co-designs Ethereum's data characteristics and an LSM-tree data structure to optimize the system I/O resource requirements. It introduces a scheme named block number related prefix hashing that makes the insertion of KV entries sequentially to mitigate the write amplification caused by compaction operations during the data synchronization process. Block-LSM claims that it can increases the synchronize throughput by 3$\times$ compared with the original design. However, the way to prefix multiple KV entries undoubtedly sacrifices extra storage space. Thus, how to minimize the need for extra space will be considered further.

\subsection{Future Optimization Directions}

This subsection lists some future research directions based on the current optimization schemes and existing defects of standard blockchain.

$\bullet$ {\it Performance Trade-off between Reads and Writes.}
The underlying storage engines employed on blockchain designs mostly use write-friendly databases (such as LevelDB), but efficient reads are also required by real applications. Therefore, the read performance should be carefully considered in further optimization. One possible direction is to improve read performance by exploring the application semantics of blockchain systems. For blockchain systems that adopt KV store as their storage engines, blockchain data are translated into KV items with a unified scheme (e.g., use the hash value of a transaction as the key and the content as the value) and stored indistinguishably.

In the access process of most blockchain systems, there are obvious semantic features. For example, querying a transaction or an account in Ethereum tends to access multiple key-value entries sequentially in the order of metadata to data, whereas these operations to retrieve key-value entries from the underlying storage are regular, that is to say, due to the one-to-one relationship between metadata and data, the next key-value pair to be accessed is confirmed after accessing a key-value pair. In persistent storage, if these data are stored sequentially, the data that is accessed continuously has a strong locality. However, most blockchain systems ignore this semantic characteristic, thus, leading to low read performance. Therefore, a future design may take advantage of the spatial locality of a blockchain system (e.g., through designing prefix-aware hash key) to balance the performance of data reads and writes.

$\bullet$ {\it Distributed Data Query.} 
Due to the increase of data amount, several blockchain systems propose to maintain their data in a distributed approach among nodes. The main problem of this distributed data storage is how to quickly locate which node contains the data required. In the future, we can apply new data structures (such as distributed hash table~\cite{maurer1975hash} and learned index~\cite{soloman2005index}) for blockchain systems to speed up the query process by heuristically locating the required data.

$\bullet$ {\it Optimization of the Underlying Storage System.} 
The key-value database has high write performance while relatively low read performance, but in the scenarios of the blockchain system, the workloads usually are read-intensive. For example, in the IOTA blockchain, when issuing a new transaction, the system needs to query all related historical transactions for unique address checking. Since the key-value database is the underlying storage system of the blockchain system, optimizing its internal strategy will not affect the efficiency of the blockchain system itself. Considering the characteristics of the blockchain system, improving the read performance at the underlying database level is also a promising and important optimization direction to increase the overall throughput of blockchain systems.

\section{Hybrid Blockchain}
\label{sec:hybrid}

\begin{table*}
   \caption{Overview of existing optimization papers for Hybrid Blockchain.}
   \label{tab:optimization for hybrid}
    \resizebox{\textwidth}{!}{
    \begin{tabular}{ccccccc}
\toprule
  Target problem      &  Solution                                                                                                  &  Authors                                                        &  Year                                               &  Weakness \\\midrule

                                &                                                                                                                &  Wood et al.~\cite{wood2016polkadot}           &  2016                                              &  No cross-chain transaction   \\

\multirow{-2}{*}{ \begin{tabular}[c]{@{}l@{}}Communication between different Blockchains\end{tabular}}                             &  \multirow{-2}{*}{ \begin{tabular}[c]{@{}l@{}}cross-chain\end{tabular}}       &  Kwon et al.~\cite{kwon2016cosmos}              &  2016                                            &  Central node exists     \\ \cline{1-5}

%Low throughput                               & Use other databases          &  Gupta et al.~\cite{DBLP:conf/vldb/Gupta20}              &  2020                                            &  Additional resource consumption     \\ \cline{1-5}

                               &                                                                                                                 &  Gupta et al.~\cite{gupta2018efficiently}                 &  2018                                      &  Increase transaction complexity \\

                             &  \multirow{-2}{*}{ \begin{tabular}[c]{@{}l@{}}Metadata model\end{tabular}} &  Gupta et al.~\cite{gupta2018building}      &  2018                                                    &  Increase transaction complexity     \\\cline{2-5}

                               &                                                                                                                 &  Wai et al.~\cite{wai2019storage}                 &  2019                                      &  Increase storage complexity \\

\multirow{-4}{*}{ \begin{tabular}[c]{@{}l@{}}Inefficient query mechanism\end{tabular}}                              &  \multirow{-2}{*}{ \begin{tabular}[c]{@{}l@{}}Use other databases\end{tabular}} &  Muzammal et al.~\cite{muzammal2018blockchain}     &  2018                       &  Increase storage complexity     \\ \cline{1-5}

                                 & data coding                                                                                                               &  Qi et al.~\cite{DBLP:conf/icde/QiZJZ20}           &  2020                                              &  Data loss   \\  \cline{2-5}

                                 &                                                                                                             &  Zheng et al.~\cite{DBLP:conf/icde/ZhengXZZYZ21}           &  2021                                              &  Decrease in decentralization   \\  

\multirow{-3}{*}{ \begin{tabular}[c]{@{}l@{}}Excessive Data Load\end{tabular}}                             &  \multirow{-2}{*}{ \begin{tabular}[c]{@{}l@{}}Data distribution\end{tabular}}       &  Bandara et al.~\cite{DBLP:journals/jsa/BandaraLFSRZ21}             &  2021                                            &  Decrease in decentralization    \\ 

\bottomrule
\end{tabular}  }         
\end{table*}

A Hybrid blockchain system maintains multiple chains in the network, and each node is only allowed to access the data in its specified chain. A hybrid blockchain usually has the same blockchain structure as a standard blockchain. To efficiently maintain multiple chains, one research direction for hybrid blockchain systems is to enable efficient communication between different chains. Moreover, similar to standard blockchain systems, another research direction is to deal with excessive data load. Compared to standard blockchain systems, hybrid blockchain systems may suffer severer excessive data load since the multi-ledger structure increases the amount of data. At the same time, it also leads to increased complexity in data processing and reduces query efficiency. Thus, next, we will introduce the optimization of hybrid blockchain systems from three aspects, namely, architecture, data structure and storage engine. Table~\ref{tab:optimization for hybrid} summarize these optimization works for the Hybrid Blockchain system.

\subsection{Optimization for Hybrid Blockchain Architecture}

Currently, one blockchain platform is usually designed for one specific application and communication is not allowed between different platforms. Cross-chain~\cite{herlihy2018atomic} is proposed by changing the hybrid blockchain architecture to enable communication between different chains. The main idea of cross-chain is called relay technology, that is, by adding a data structure to the two separate chains so that they can interact with each other through the data structure. Taking digital assets as an example, if the barriers between different chains can be broken, cross-chain transactions of various digital assets can be realized. Two most influential cross-chain technologies are Polkadot~\cite{wood2016polkadot} and Cosmos~\cite{kwon2016cosmos}.  Cosmos targets cross-chain digital asset transactions, while Polkadot focuses on general cross-chain communication.

%Polkadot~\cite{wood2016polkadot} believes that blockchains should not solve all problems as they do now but should solve each specific problem within one chain. In this case, communications between different chains are needed, and 

In Polkadot~\cite{wood2016polkadot}, to support communication, a security mechanism is needed to make mutual trust between each other. Polkadot is the protocol that makes it possible to communicate with other chains and exchange data or assets (tokens) on the basis of maintaining security. Polkadot's backbone network is called the relay chain. It mainly realizes the interconnection with various parachains based on Ethereum. Each secondary chain is a separate blockchain network. Polkadot also aims to upgrade other public chains so that Ethereum can communicate directly with other chains. 
However, for practical applications, only support communication between different chains may not be enough, which further incurs the design of the cross-chain transaction.

Cosmos~\cite{kwon2016cosmos} is a cross-chain network designed to solve cryptocurrency problems, such as lack of scalability. By using the Cosmos network, users can transfer tokens from one chain to another through internal cross-chain communication. The Cosmos network is composed of various blockchains, which are called zones. Each zone contains a Byzantine fault-tolerant consensus mechanism like Tendermint. The first blockchain in Cosmos will be called the Cosmos hub, which is a multi-asset blockchain used as a central hub. Cosmos realizes the interconnection between different regions through the inter-blockchain communication (IBC) protocol running on the backbone network cosmos Hub.
In Cosmos, cross-chain transactions need to be implemented through a central hub, which is similar to the existence of a central node. Once it is attached and performs some malicious behaviors, the security of the entire blockchain system cannot be guaranteed.

%{\color{red} this paragraph is strange here. I didnot get its relation with other part and the description of the first sentence is also not clear, should we eliminate it?}
%\textcolor{red}{
%The low throughputs of existing permissioned fabrics are due to missed opportunities during their design and implementation. Gupta et al.~\cite{DBLP:conf/vldb/Gupta20} propose ResilientDB which lays down an efficient client-server architecture. At the application layer, multiple clients are allowed to co-exist, each of which creates its own requests, clients and replicas use the transport layer to exchange messages across the network. ResilientDB also provides a storage layer where all the metadata corresponding to a request and the blockchain is stored. For each replica, there is an execution layer where the underlying consensus protocol is running on the client request, and the request is executed.
%}

\subsection{Optimization for Hybrid Blockchain Data Structure}
As mentioned, hybrid blockchain systems have the same data structures as standard blockchain systems. Therefore, the technologies on data structure optimization for standard blockchain systems are also applicable to hybrid blockchain systems. Moreover, several studies~\cite{gupta2018efficiently, gupta2018building} are proposed specifically to improve the hybrid blockchain performance.

%处理时间查询的问题
Gupta et al.~\cite{gupta2018efficiently} discuss the problem of how to efficiently handle temporal queries on Hyperledger Fabric. The temporal nature of the data inserted by the Hyperledger Fabric transactions can be leveraged to support various use cases. They present two models to overcome these limitations and improve the performance of temporal queries on Fabric. The first model creates a copy of each event inserted by a Fabric transaction and stores temporally close to events together on Fabric. The second model keeps the event count intact but tags some metadata to each event inserted on Fabric, and temporally close events share the same metadata.

Further, Gupta et al.~\cite{gupta2018building} present a variant based on these two models to better handle skew data patterns. The variant significantly outperforms the approaches presented in ~\cite{gupta2018efficiently} when Fabric contains skew data. They also discuss the performance tradeoffs among the variant across various dimensions such as data storage, query performance, event insertion time, etc.
Same as the Standard Blockchain system, improving query efficiency also increases the complexity of the entire system model, requiring more resources to complete the entire transaction processing.

\subsection{Optimization for Hybrid Blockchain Storage Engine}

\subsubsection{Optimization for Excessive Data Load}
Like the standard blockchain, some research papers aim to optimize the excessive data storage problem for hybrid blockchain. They also adopt data sharding and data coding methods. Some up-mentioned methods for the standard blockchain method are still applicable to the hybrid blockchain. Further, there are some other designs for the special application execution environment of the hybrid blockchain.

Qi et al.~\cite{DBLP:conf/icde/QiZJZ20} use erasure coding to reduce the storage pressure of the permissioned blockchain. Specifically, two ideas are designed to achieve scalability and speed up reading performance: i) a four-phase re-encoding protocol based on PBFT to promise the availability of all blocks; ii) a multiple replication manner to ensure efficient access of blocks.

Zheng et al.~\cite{DBLP:conf/icde/ZhengXZZYZ21} propose Meepo to use multiple execution environments per organization. Meepo also uses the idea of data sharding which includes two main processes: cross-epoch and cross-call. After the consensus of each block ends, each shard begins to communicate across shards. This process is so-called cross-epoch. While cross-call is sharing the message from one shard to another, resulting from cross-shard transactions, such as cross-shard payment. Meepo aggregates cross-calls into several cross-epochs in order and proposes a partial cross-call merging strategy to improve the smart contract flexibility in sharding environments.

Bandara et al.~\cite{DBLP:journals/jsa/BandaraLFSRZ21} propose Rahasak, in which all blocks, transactions and asset information are stored in a distributed database. Every blockchain peer comes with a distributed database node; these nodes are connected as a ring cluster. After executing a transaction, state update in a peer is distributed and replicated with other peers via underlying distributed databases’ sharding algorithm. However, this approach causes the loss of the immutability of the blockchain itself and further incurs more security problems.

\subsubsection{Optimization for Query Engine}
To improve the query efficiency of hybrid blockchain systems, Wai et al.~\cite{wai2019storage}  and Muzammal et al.~\cite{muzammal2018blockchain} propose to optimize the hybrid blockchain storage engine. 

 Wai et al.~\cite{wai2019storage}  first analyze the factors to determine the block size. Based on the analysis, they change the block size to reduce the waiting time of a transaction before creating a block. Therefore, using a proper block size can finally increase the throughput of transactions in data writing. Moreover, the MongoDB database is used as an off-chain to duplicate the metadata of the block. This system can support temporal queries of transactions on a blockchain. The access time of the proposed system can be shortened compared to Hyperledger Fabric.

%ChainSQL
Muzammal et al.~\cite{muzammal2018blockchain} implement a protocol named ChainSQL that combines Ripple and distributed databases between all participants in the Byzantine environment using blockchains. The protocol can process queries in RDBMS after all transactions are transferred. The transactions are stored in the blockchain, whereas the actual data is stored in the database. Thus, ChainSQL not only provides the instantaneity of the traditional databases but also enhances the security of the blockchain.
Similarly, storing metadata in a dedicated database also increases storage complexity and additional resource consumption.

\subsection{Future Optimization Directions}

For hybrid blockchain systems, the key challenge is how to coordinate the cross-chain data exchange and how to provide efficient queries.

With the development of network topology structure, a hybrid blockchain system may contain hundreds to thousands of sub-chains, thus the verification delays of cross-chain transactions increase exponentially. 
To alleviate this issue, one potential direction is to combine network sharding technology with cross-chain technology, that is, dividing nodes in the network into different shards. The nodes in each shard only process the transactions within a part of the chains or are initiated by these chains to reduce network load.

On the other hand, the most popular hybrid blockchain transaction technologies are the notary mechanism and chain relay technology, both of which introduce a third party for verification. Therefore, for the query process, a possible improvement direction is to establish a new auxiliary query module in the third party by establishing a SQL-friendly database (e.g., SQLite) to record each transaction and directly retrieve data from the third party when querying.

\section{DAG-based Blockchain}
\label{sec:dag}

\begin{table*}
   \caption{Overview of existing optimization papers for DAG-based Blockchain.}
   \label{tab:optimization for dag}
    \resizebox{\textwidth}{!}{
    \begin{tabular}{ccccccc}
\toprule
  Target problem      &  Solution                                                      &  Authors                                                        &  Year                                           &  Weakness \\\midrule  
                                & Use NVM                                                    &  Wang et al.~\cite{wang2019re}                     &  2019                                           &  Not really deployed   \\  
\multirow{-2}{*}{ \begin{tabular}[c]{@{}l@{}}Transaction execution speed\end{tabular}}                               &  Change transaction verification mode         &  Ferraro et al.~\cite{ferraro2018iota}              &  2018                                            &  Reduce transaction security     \\ \cline{1-5}

                                &                                                     &  Li et al.~\cite{DBLP:journals/corr/abs-1805-03870}                     &  2018                                           &  Unable to resolve malicious transactions   \\  
\multirow{-2}{*}{ \begin{tabular}[c]{@{}l@{}}Low throughput\end{tabular}}                               &  \multirow{-2}{*}{ \begin{tabular}[c]{@{}l@{}}DAG-Based architecture\end{tabular}}         & Li et al.~\cite{DBLP:conf/usenix/LiLZYWYXLY20}              &  2020                                            &  Unable to resolve malicious transactions    \\ \cline{1-5}

                               & Cuckoo filter                                               &  Shaffq et al.~\cite{shafeeq2019curbing}        &  2019                                           &  Hard to be deployed \\         
                               &  Two-level bloom filter                                &  Wang et al.~\cite{wangtianyu}                    &  2020                                           &  Ignores the temporal locality of the addresses     \\
\multirow{-3}{*}{ \begin{tabular}[c]{@{}l@{}}Unique address checking\end{tabular}}                               &  Hotness aware bloom filter                        &  Zhu et al.~\cite{zhuwenbin}                      &  2020                                           &  Low safety and fault tolerance     \\
\bottomrule
\end{tabular}  }         
\end{table*}

DAG-based blockchains are proposed for IoT environments to provide thousands of transactions per second (TPS). Existing techniques mainly focus on optimizing transaction verification and alleviating the storage burden caused by the unique address checking process. Table~\ref{tab:optimization for dag} summarize these optimization works for the DAG-Based Blockchain system.

\subsection{Optimization for DAG-based Blockchain Architecture}

Wang et al.~\cite{wang2019re} present Re-Tangle, a novel DAG-based blockchain acceleration architecture that explores the opportunity of performing massive parallel operations with low energy cost. Specifically, ReRAM (Resistive Random Access Memory), an emerging non-volatile memory with computation capacity, is utilized to optimize the Tangle working process for DAG-based blockchain systems. Re-Tangle consists of a random walking module and a transaction validation module, which transfers Tangle functions into ReRAM-based logic analog computation units. In the random walking module, Re-Tangle maintains an exponentiation translator to reduce its design complexity and improves its computation efficiency for exponentiation calculation. In the transaction validation module, Re-Tangle leverages a highly parallel modular unit to accelerate the validation of different tags in a transaction. 
Although the use of a ReRAM-based accelerator can significantly improve overall transaction speed and reduce resource consumption, it has not really landed yet, so currently the solution is practical.

Ferraro et al.~\cite{ferraro2018iota} propose a simple modification to the attachment mechanism for the Tangle (the IOTA DAG architecture). This modification ensures that all transactions are validated in a finite time and preserves essential features of the popular Monte-Carlo selection algorithm. They propose a new hybrid tip selection algorithm that resolves the dichotomy of double-spend avoidance and orphans and results in a Tangle where all transactions are validated in a finite time.
The improved algorithm increases the coming speed of transactions into the chain but reduces the security of the original algorithm to a certain extent, which may lead to the phenomenon of malicious winding into the chain, such as double-spending.

For the standard blockchain, the solution adopted for concurrent block chaining is to select the longest chain. Since the transactions in the short chain are discarded, the overall throughput is reduced. To address this, Conflux~\cite{DBLP:journals/corr/abs-1805-03870} organizes blocks into a novel Tree-Graph structure, which is a tree embedded inside a direct acyclic graph (DAG). Conflux assigns a weight to each block, which indicates the amount of finality that the block contributes to its ancestors. To further adjust to environmental changes, Li et al.\cite{DBLP:conf/usenix/LiLZYWYXLY20} redesign the consensus protocol of Conflux that encodes two different block generation strategies: an optimistic strategy that allows fast confirmation and a conservative strategy that ensures the consensus progress. Conflux uses its novel adaptive weight mechanism to combine these two strategies into a unified consensus protocol to cope with different workloads.

\subsection{Optimization for data storage engine}
Unique address checking is one essential process to guarantee that each transaction from each user account (wallet) is associated with a unique private/public key pair, in which the public and private keys will be utilized as the address and for the signature of the transaction, respectively. Currently, DAG-based blockchain systems usually adopt a database-based approach that relies on the time and space consuming database query process to perform the unique address checking process. To address this problem, several approaches have been proposed.

Shafeeq et al.~\cite{shafeeq2019curbing} design a cuckoo filter embedded into IOTA lightweight clients to help accelerate this address checking process. This design is based on the assumption that the addresses can be reused after a snapshot, which is performed periodically to reduce Tangle size. They build a Bloom filter for addresses in the current Tangle to promise uniqueness. However, this assumption is unrealistic and makes this design hard to be deployed in practice.

Wang et al.~\cite{wangtianyu} propose a method called ABACUS to utilize a two-level partitioned bloom filter to perform address checking. Partitioned bloom filters are responsible for storing all spent addresses according to their prefixes, and one SBF (Sub Bloom Filter) of each partitioned bloom filter is kept in memory as a write buffer. However, this approach focuses on optimizing bloom filter update operations so as to speed up unique address checking. Besides, ABACUS assumes all the address checking processes are performed within the paying account and ignores the temporal locality of the addresses.

Zhu et al.~\cite{zhuwenbin} propose a Hotness aware and Fine-grained Bloom Filter (HF-BF) for unique address checking in DAG-based blockchains. They divide the whole address space into several subspaces according to the address prefix. For each subspace, a Bloom Filter Group (BFG) that consists of u Bloom Filter units (BFUs) corresponding with K$_u$ hash functions is used to perform the unique address checking. A BFU is a fine-grained bloom filter composed of a bit array for K$_u$ hash functions. To manage the large storage footprint of BFUs and decrease the number of I/Os for checking addresses as much as possible, they design an adaptive BFU management scheme to schedule active BFUs between memory and disk.
This solution has certain security risks. Since the address is not stored, once an error occurs in the blockchain network, the transaction cannot be traced back.

%They propose three management schedules: BFG-LRU, BFU-LRU, and Analytical adjuster, to perform the checking process in memory as much as possible.

\subsection{Future Optimization Directions}

In DAG-based blockchain systems, a full node is responsible for making consensus, adding and confirming all transactions, thus consuming lots of computation and memory resources, and making it become the system bottleneck. 
As the number of transactions increases in a blockchain system, a full node may crash under heavy workloads. Thus, how to mitigate the computation and memory pressure of a full node in a DAG-based blockchain system is an interesting and crucial problem.

Unique address checking is one key process to protect users' security and privacy when transactions are carried out in DAG-based blockchain systems. To guarantee address uniqueness, the address size must be large enough, and with the increment of the number of transactions, it becomes challenging to efficiently perform unique address checking. Several studies propose to adopt bloom filters to speed up the query process, but it still causes many I/O operations. One future optimization direction is to reuse addresses to mitigate resource consumption caused by the unique address checking process.
For example, we can use an additional signature scheme (e.g., Merkle OTS signature algorithm) to support reusable addresses based on the current WOTS scheme. Another possible direction is to leverage the snapshot mechanism of DAG-based blockchain systems, that is, unique addresses are only required to be guaranteed in one snapshot period. After each snapshot, addresses with empty balances can be reattached to the DAG structure for address reusing.

\section{Blockchain-assisted Databases}
\label{sec:app}

%table4
   \begin{table*}
   \caption{A brief overview of existing works of using blockchain-assist database.}
    \label{tab:application}
    \begin{tabular}{ccccc}
\toprule
  Target problem                                                      &  Authors                                                                          &  Year                         &  Weakness                                                                                                    \\\midrule

                                                             &  AliNSF et al.~\cite{DBLP:conf/usenix/AliNSF16}                            &  2016                         &   Excessive storage consumption                                       \\ 
                                                             &  Grabis et al.~\cite{DBLP:conf/icde/GrabisSZ20}                            &  2020                         &   Excessive storage consumption                                       \\ 
                                                             &  Yang et al.~\cite{DBLP:journals/jsa/YangLW18}                            &  2018                         &   Excessive storage consumption                                       \\ 
                                                            &  Konsta et al.~\cite{DBLP:conf/icde/KonstaMDNK21}                           &  2021                         &   Excessive storage consumption                                       \\ 
                                                             &  Ali et al.~\cite{DBLP:journals/jsa/AliGAL19}                            &  2019                         &   Excessive storage consumption                                       \\ 
                                                      &  Eltayieb et al.~\cite{DBLP:journals/jsa/EltayiebEHL20}                            &  2020                         &   Excessive storage consumption                                       \\ 
                                                      &  Qin et al.~\cite{DBLP:journals/jsa/QinHYL21}                            &  2021                         &   Excessive storage consumption                                       \\ 
                                           &  Chen et al.~\cite{chen2017improved}                            &  2017                         &  Excessive storage consumption                                                                                \\ 
 \multirow{-9}{*}{ Data security protection.}            & El-Hindi et al.~\cite{el2019blockchaindb}                         &  2019                &   Excessive storage consumption                                         \\ \cline{1-4}

                                                                 &  McConaghy et al.~\cite{mcconaghy2016bigchaindb}     &  2016                        &  Inefficient traceability                                                                                   \\ 
     \multirow{-2}{*}{ Scalability}              &  Li et al.~\cite{li2018block}                                               &  2018                         &  Inefficient traceability                                                                                 \\ 
\cline{1-4}

                                                                               &Li et al.~\cite{li2017searchable}                                        &  2017                          &  \begin{tabular}[c]{@{}l@{}}Cannot available \\ for the dynamic data\end{tabular}                                                           \\  
                                                                               &Wang et al.~\cite{wang2018blockchain}                            &2018                          &  Reduce the scalability                                                                         \\
  \multirow{-4}{*}{Secure Query}                               &  Do et al.~\cite{do2017blockchain}                                   &  2017                         &  Excessive network load                                                                                   \\ \cline{1-4}

                                                                                &  Wilkinson et al.~\cite{wilkinson2014metadisk}               &  2014                        &  Excessive network load                                                                                 \\  
                                                                                 & Yang et al.~\cite{yang2018blockchain}                          &2018                           & Excessive network load                                                                                 \\
  \multirow{-3}{*}{Authentication}                               &  Ali et al.~\cite{ali2018blockchain}                                   &  2018                        &  Complex query process                                                                                 \\ \bottomrule
\end{tabular}          
\end{table*}

With decentralization, non-tampering, and data consistency, blockchain systems are proposed to assist the design of databases from four aspects: data security protection, scalability, secure query, and authentication. Table ~\ref{tab:application} provides a summary of these studies.

$\bullet$ {\it Data security protection.}
Benefiting from its safety and immutability, many works choose to use the blockchain as the underlying storage to solve the data security problem. In the blockchain environment, data can be shared safely, and the built-in consensus mechanism ensure that the use of data is subject to certain rules.

Chen et al.~\cite{chen2017improved} propose an improved P2P file database based on IPFS blockchain. They address the low-throughput problem for individual users in IPFS by introducing the role of content service providers. Considering data reliability and availability, storage overhead, and other issues for service providers, they provide a novel zigzag-based storage model to improve the throughput of the block storage model. They chose Bitcoin as the underlying blockchain at the bottom layer of the storage model. All transactions are processed in the upper layer, and then are sent and stored to the blockchain layer. Verified transactions are added to the underlying blockchain, while unverified transactions are discarded.
In this environment, the blockchain system is only used as storage. Once the verification service is completed by the node as the service provider, the centralization and non-tampering characteristics of the blockchain system are erased, and the overall security of the system cannot be guaranteed.

El-Hindi et al.~\cite{el2019blockchaindb} utilize blockchains as a storage layer and introduce a database layer on top to extend blockchains by classical data management techniques (e.g., sharding). Further, BlockchainDB provides a standardized key/value-based query interface to facilitate the adoption of blockchains for data sharing. Thus, by using BlockchainDB, we can not only improve the performance and scalability of blockchains for data sharing but also decrease the implementation complexity. 
On the other hand, the database layer used for expansion will bring additional storage resource consumption.

AliNSF et al.~\cite{DBLP:conf/usenix/AliNSF16} propose the design and implementation of a new blockchain-based naming and storage system, called Blockstack. Unlike previous blockchain-based systems, Blockstack separates its control and data planes: it keeps only minimal metadata (namely, data hashes and state transitions) in the blockchain and uses external data stores for actual bulk storage. Grabis et al.~\cite{DBLP:conf/icde/GrabisSZ20} elaborate a method for efficient distributed storage and sharing of personal data assets within a community of users. The access to these data is controlled using a blockchain and smart contracts that define access conditions. Yang et al.~\cite{DBLP:journals/jsa/YangLW18} propose the smart toy edge computing-oriented data exchange accounting system based on the blockchain. Through smart contracts, each node in a data-exchange P2P network makes consensus and endorsements for each other, queries the transaction record and checks the payment bill from its local blockchain peer. Same as~\cite{DBLP:journals/jsa/YangLW18}, there are also some other works~\cite{DBLP:conf/icde/KonstaMDNK21,DBLP:journals/jsa/AliGAL19,DBLP:journals/jsa/EltayiebEHL20,DBLP:journals/jsa/QinHYL21} that use the storage of data on the blockchain and the use of smart contracts for secure cloud data sharing.

$\bullet$ {\it Scalability.}
Several studies adopt blockchain features to distributed databases for scalability improvement~\cite{mcconaghy2016bigchaindb, li2018block}. For example, BigchainDB~\cite{mcconaghy2016bigchaindb} leverages the blockchain characteristics like decentralized control, immutability, creation, and movement of digital assets on a distributed database. BigchainDB is jointly constructed by database nodes (such as MongoDB instances) available to the enterprise. These nodes store immutable information about assets in a synchronized manner. By implementing Tendermint's BFT consensus algorithm~\cite{buchman2018latest}, the data stored in the network is propagated and synchronized among all nodes, ensuring the integrity of the network. Li et al.~\cite{li2018block} propose a blockchain-based security architecture for distributed cloud storage, where users can divide their files into encrypted data chunks and upload those data chunks randomly onto the P2P network nodes that provide free storage capacity. They customize a genetic algorithm to allocate a file block replica between users and data centers in the distributed cloud storage environment.
With the equipment of blockchain features, the confidentiality of distributed storage has been enhanced. However, this method is only suitable for data environments that cannot be changed anymore. At the same time, the file is divided into data blocks, resulting in a stronger data dispersion, making the query more difficult.

$\bullet$ {\it Secure Query. }
Several studies utilize blockchains to ensure secure query of encrypted data in databases~\cite{li2017searchable,do2017blockchain}. Li et al.~\cite{li2017searchable} combine blockchains with SSE (Searchable Symmetric Encryption) to store encrypted data in a decentralized manner. To support data retrieval, they innovatively construct an SSE model using blockchains (SSE-using-BC) to ensure data privacy. They store an encrypted file and its index as a transaction on a blockchain for subsequent queries.

Do et al.~\cite{do2017blockchain} introduce a system that leverages blockchain technology to provide secure data storage with keyword search service. The system allows clients to upload their data in an encrypted form. Moreover, the system can distribute the data content to cloud nodes and ensure data availability using cryptographic techniques. In the system, data owners can grant permission for others to search on the data. Finally, the system supports private keyword search over encrypted datasets.

Wang et al.~\cite{wang2018blockchain} study the data storage and sharing scheme for decentralized storage systems and propose a framework that combines the decentralized storage system interplanetary file system, the Ethereum blockchain, and attribute-based encryption (ABE) technology. Based on the smart contracts of Ethereum, the keyword search function on the ciphertext of the decentralized storage systems is implemented to solve the problem that the cloud server may not return all searched results or return wrong results in the traditional cloud storage systems. It is relatively novel to provide search services based on smart contracts, but at the same time, searching will consume a lot of storage resources and reduce the scalability of the entire system.

$\bullet$ {\it Authentication.}
Several studies have put forward the concept of verifiable databases, in which the main idea is to allow databases to be verified and shared. Several works~\cite{wilkinson2014metadisk,ali2018blockchain} propose to implement this concept by storing data in traditional databases and storing the digest of the data in an underlying blockchain. Wilkinson et al.~\cite{wilkinson2014metadisk} devise a new model in which a blockchain system serves as the backbone for a distributed application, Metadisk. This application operates autonomously as a peer-to-peer network of nodes running open-source code. A cryptocurrency will serve as both an incentive and payment mechanism, while a separate blockchain will be used as a datastore for file metadata.  Ali et al.~\cite{ali2018blockchain} design a blockchain-based data storage framework for PingER (a worldwide end-to-end Internet performance measurement project) to remove its total dependence on a centralized repository. They adopt the permissioned blockchain and Distributed Hash Tables (DHT) for this purpose. In the proposed framework, file metadata are stored on the blockchain whereas file contents are stored off-chain through DHT at multiple locations using a peer-to-peer network with PingER monitoring agents. This design provides decentralized storage, distributed processing, and efficient lookup capabilities to the PingER framework.

Yang et al.~\cite{yang2018blockchain} propose a novel blockchain-based data deletion scheme, which can make the deletion operation more transparent. The data owner can verify the deletion result no matter how malevolently the cloud server behaves. Besides, with the application of blockchain, the proposed scheme can achieve public verification without any verification from a trusted third party.
Unlike~\cite{wilkinson2014metadisk} and~\cite{ali2018blockchain}, the blockchain system is not used to store metadata, but stores the records of data deletion behavior, which provides sufficient security. However, the problem of excessive network load still exists.

\section{Related Work}
\label{sec:Related}

Previous survey papers for blockchain systems mainly focus on system architecture, security challenges, consensus mechanisms, and applications. However,  rare survey work investigates the underlying data management system of blockchain systems.

$\bullet$ {\it System architecture.}
Zheng et al.~\cite{zheng2017overview} provide a comprehensive overview of blockchain technology, including blockchain architectures and typical consensus algorithms, and list the technical challenges, the latest developments, and possible future trends. Dai et al.~\cite{dai2019blockchain} introduce the integration of blockchain and the Internet of Things, propose related architecture recommendations, and outline the open research directions in this promising field. 

$\bullet$ {\it Security challenges.}
Feng et al.~\cite{feng2019survey} analyze the privacy threats in blockchain systems and discuss the existing cryptographic defense mechanisms, namely anonymity and transaction privacy protection. Joshi et al.~\cite{joshi2018survey} conduct a comprehensive investigation of blockchain technology by discussing its structure, different consensus algorithms, and the challenges and opportunities faced by data security and privacy in blockchains.

%consensus
$\bullet$ {\it Consensus mechanisms.}
Wang et al.~\cite{wang2019survey} conduct in-depth research on the latest consensus protocols and discuss several unresolved problems and future research directions. Nguyen et al.~\cite{nguyen2018survey} review the blockchain consensus algorithms and introduce typical applications. 

%application
$\bullet$ {\it Applications.}
Lao et al.~\cite{lao2020survey} conduct a systematic survey on the key components of an IoT blockchain and study popular blockchain applications. Duy et al.~\cite{duy2018survey} discuss the impact of blockchain technology, as well as the opportunities and challenges of using it in real-world scenarios. Tasatanattakool et al.~\cite{tasatanattakool2018blockchain} investigate the blockchain applications using blockchain technology and the challenges they face. Zhang et al.~\cite{zhang2020blockchain} study the application of blockchain traceability technology in various fields, the decentralized application of blockchains and the application of other blockchains in data security protection. Al-Jaroodi et al.~\cite{al2019blockchain} review different industrial applications that have adopted blockchain technologies, and discuss the opportunities, benefits and challenges of incorporating blockchains into different industries.

\section{Conclusion}
\label{sec:Conclusion}

Blockchain with its key features - decentralization, persistence, anonymity, and auditability - shows great promise. In this paper, we provide a comprehensive overview of the underlying data management techniques of blockchain systems. We first divide the blockchain data management techniques into four layers, namely, blockchain architecture, blockchain data structure, blockchain storage engine, and blockchain-assisted database for three typical kinds of blockchain systems: standard blockchain, hybrid blockchain, and DAG-based blockchain.
We then summarize the existing solutions and ideas that focus on optimizing the blockchain architecture, blockchain data structure, and blockchain storage engine.
Finally, we give an overview of existing works of using the blockchain-assisted database. With the continuous emergence of applications based on blockchain technology, we plan to conduct in-depth investigations in the future to improve the operational efficiency and query efficiency of blockchain systems.

\bibliographystyle{ACM-Reference-Format}
\bibliography{sample-acmsmall}

\end{document}